\documentclass[11pt]{article}
\usepackage{moriond,epsfig}

\def\be{\begin{equation}}
\def\ee{\end{equation}}
\def\bea{\begin{eqnarray}}
\def\eea{\end{eqnarray}}

\begin{document}
\vspace*{4cm}
\title{DARK MATTER EXPERIMENTS AT BOULBY MINE}

\author{V. A. KUDRYAVTSEV}

\address{Department of Physics and Astronomy, University of Sheffield,
Sheffield S3 7RH, UK\\
\vspace{0.2cm}
on behalf of the Boulby Dark Matter Collaboration\\
(Sheffield, RAL, Imperial College, Edinburgh, UCLA, Temple, Occidental, 
Texas A\&M, ITEP and Coimbra)}

\maketitle\abstracts{
The Boulby Dark Matter Collaboration (BDMC) is running several 
experiments for particle (WIMP) dark 
matter search in the Underground Boulby Laboratory. These
include a liquid xenon
detector (ZEPLIN I) and a low pressure gas TPC with directional
sensitivity (DRIFT I). Next stage double-phase xenon detectors 
ZEPLIN II and ZEPLIN III, and a new TPC DRIFT II will be
installed at Boulby in a few months. Recent results from the running
experiments are discussed and future programme towards
large-scale detectors is presented.
}

\vspace{-0.3cm}
\section{Introduction}
It is believed that 20\% of the Universe may consist of non-barionic 
dark matter. 
Supersymmetric theories provide a good candidate -- neutralino or 
Weakly Interacting Massive Particle (WIMP).
Due to very small cross-section of WIMP-nucleus interactions,
very sensitive and massive detectors are required to detect WIMPs.
There are three key requirements for direct dark
matter detection technology: (1) low intrinsic radioactive background
from detector and surrounding components;
(2) good discrimination between electron recoils produced by remaining
gamma background and nuclear recoils expected from WIMP interactions; 
(3) low energy threshold to achieve
maximal sensitivity to WIMP-induced nuclear recoils. 

The UK Dark Matter Collaboration has been running a Dark Matter 
programme at Boulby Mine (North Yorkshire, UK) at a vertical depth 
of 2800 m w.\,e. for more than a decade.
Three major programmes were pursued so far: 
i) NAIAD experiment (an array of NaI(Tl) crystals) is almost completed and
the resources have been moved to other projects, more sensitive
to the particle non-baryonic dark matter;
ii) detectors based on liquid xenon, 
which have high background discrimination power, have been 
developed and are either running or being commissioned, 
iii) a low
pressure gas Time Projection Chamber (TPC) with a potential of
directional sensitivity has been constructed and is operating at Boulby.
Two later projects are carried out in collaboration with international 
groups from Europe and USA.

\vspace{-0.2cm}
\section{Liquid xenon experiments}

Nuclear recoil discrimination in liquid
xenon is feasible by measuring both the scintillation light and the
ionisation produced during an interaction, either directly or through
secondary recombination. Meanwhile, the chemical inertness and
isotopic composition of liquid xenon provide intrinsically high
purity and routes, in principle, to further purification using
various techniques. The heavy nuclei of
xenon also have the advantage of providing a large spin-independent
coupling.

Any recoil in liquid Xe gives rise to both ionisation and excitation of Xe atoms.
The de-excitation result in the emission of 175 nm photons from either 
singlet (with decay time $\sim 3$ ns) or triplet ($\sim 27$ ns) states.
The ratio single/triplet is several times higher for nuclear recoils compared 
to electron recoils.
In the absence of an electric field, the ions recombine with electrons to 
produce 
excited Xe atoms again. The recombination time depends on 
the ionisation density: for nuclear recoils, 
it is very high and the recombination is very fast.
For electron recoils, the lower density leads to longer times.

The ZEPLIN I detector (ZonEd Proportional scintillation in LIquid Noble 
gases -- shown in Figure~1)
consisted of liquid Xe with 3.2 kg fiducial mass incased in a 
copper vessel
and viewed by 3 PMTs through silica windows. 
PMT signals were digitised using a digital oscilloscope driven by a
Labview based software at the beginning of the experiment or using an Acqiris
CompactPCI based DAQ system later on. The detector itself was
enclosed in a 0.93 tonne active scintillator veto, its function being to veto
gamma events from the PMTs and the surroundings. 

The detector was triggered by a 3-fold coincidence of single 
photoelectron pulses in each tube.
With a light yield of at least 1.5 photoelectrons/keV in the data runs, 
this gave a 2 keV threshold.
The trigger efficiency was calculated using Poissonian statistics.
Daily energy calibration was performed with a $^{57}$Co source 
automatically placed between target and veto.
The 122 keV $\gamma$s were absorbed within $\approx$3 mm of their path
in the bottom part of the target, 
making it a calibration point source.
A 30 keV K-shell X-ray was also observed in the spectrum, 
its presence was confirmed through the GEANT4 simulation.
A full light collection simulation was performed, 
showing non-uniform of light collection efficiency. 
This affected the measured energy of an event and was 
observed in higher energy gamma calibrations ($^{60}$Co, $^{137}$Cs sources): 
as different parts of the target were illuminated, the peak position 
was shifted, which reflected the reduction in light yield. 
The observations matched well the light collection efficiency simulation.

\begin{figure}[htb]
\begin{center}
\includegraphics[width=5.5cm,height=5.5cm]{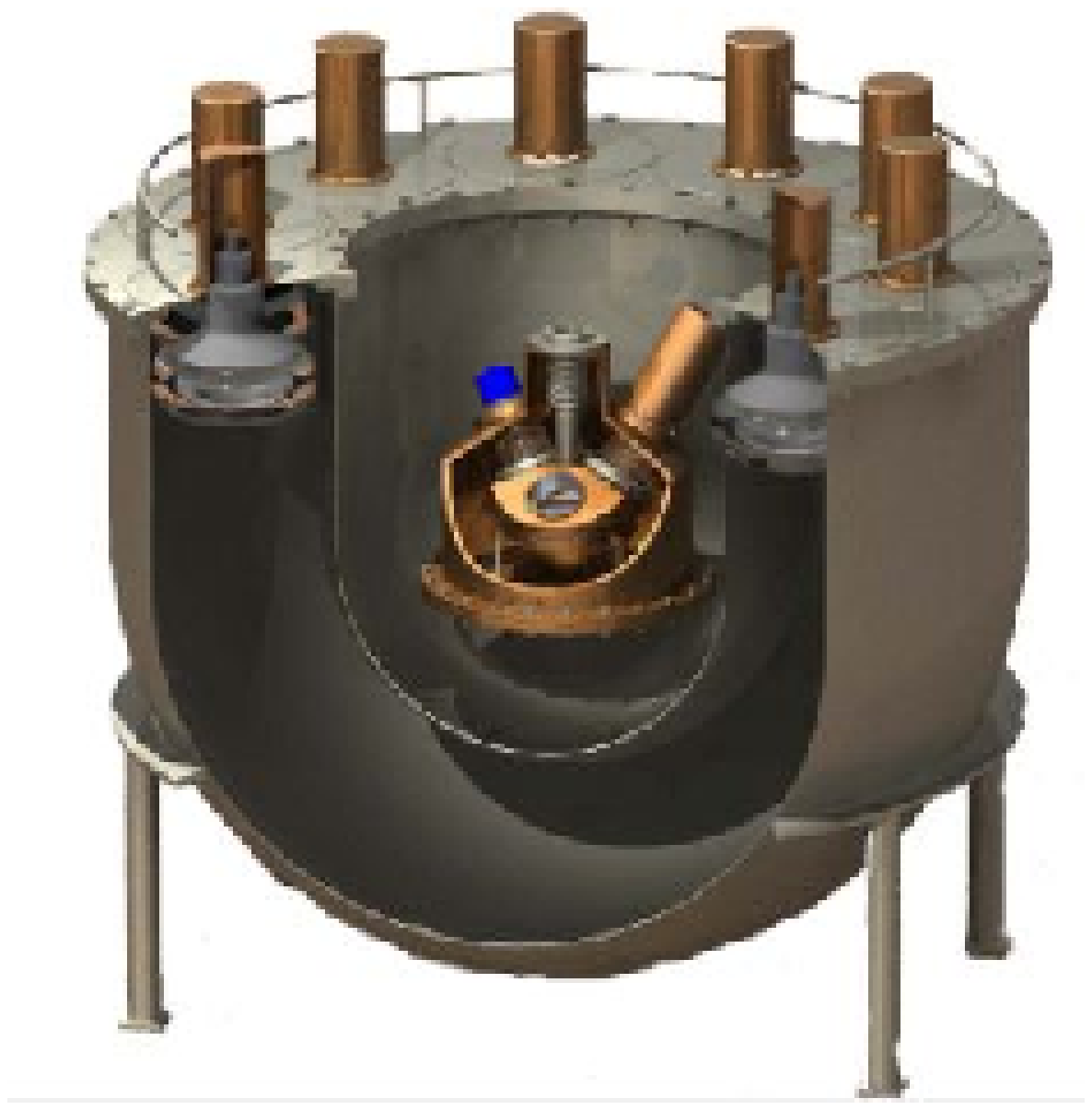}
\label{z1}
\hspace{0.5cm}\includegraphics[width=8.5cm,height=5.5cm]{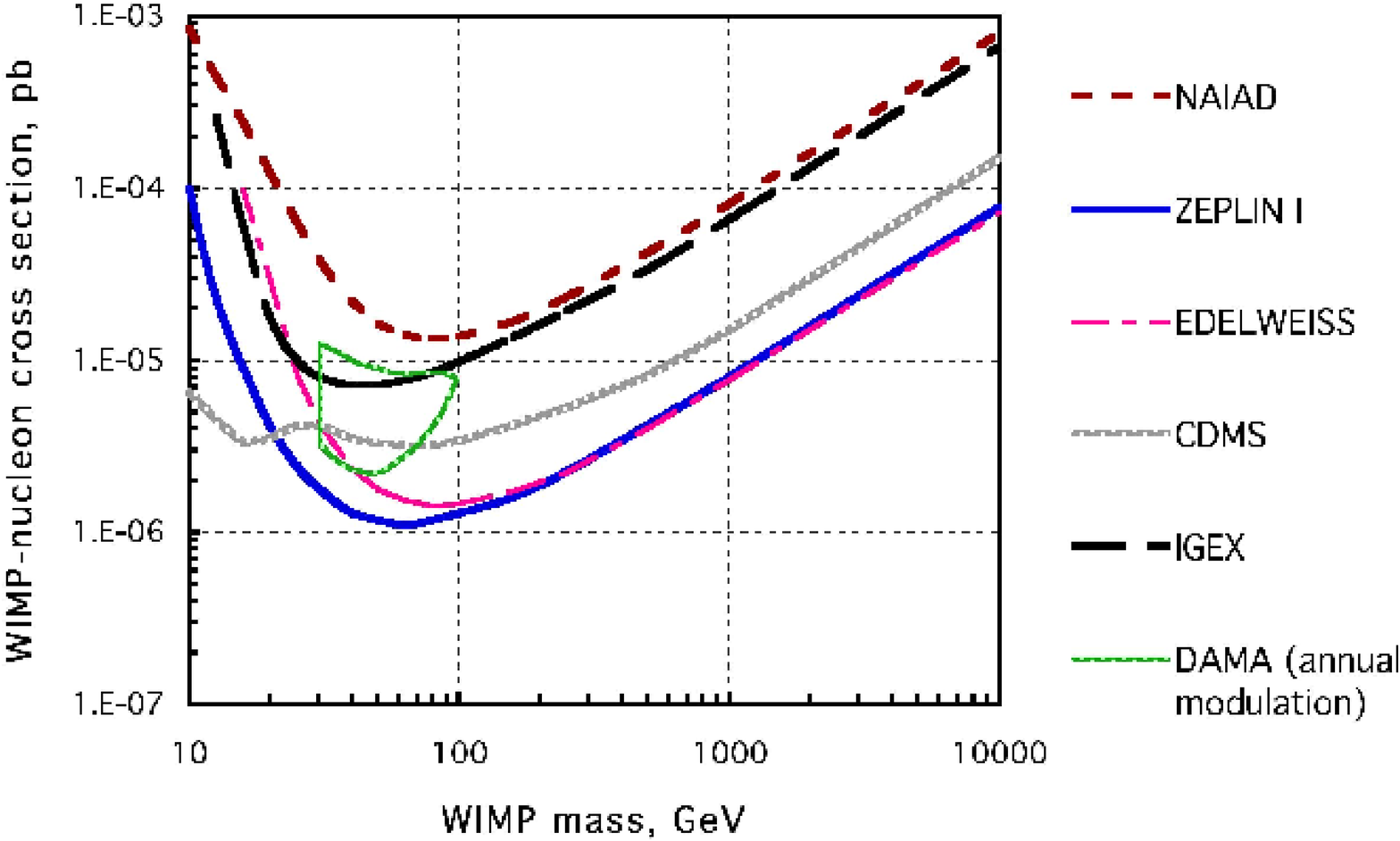}
\label{limits}
\parbox{5.5cm}{Figure 1: Computer view of the ZEPLIN I experiment 
showing the target vessel with three PMTs encased in an active veto 
system.}
\hspace{0.5cm}
\parbox{9.0cm}{Figure 2: Preliminary limits on WIMP-nucleon spin-independent
cross-section from the NAIAD and ZEPLIN I experiments are shown together
with some other limits and with a region of parameter space
consistent with the DAMA signal (see \cite{nigel} for 
recent limits).}
\end{center}
\end{figure}

ZEPLIN I had better
sensitivity than NaI detectors due to its improved discrimination at low
energies.
Background discrimination was possible due to the difference 
in the characteristic time between nuclear and electron recoil pulses. 
Our standard procedure of data analysis involved calculation of the mean time for
each scintillation pulse. The mean times of these
pulses followed a gamma density distribution with characteristic times, 
which were 
very much different (by about a factor of 2 at low energies) 
for electron and nuclear recoils. Electron and nuclear recoils thus gave
rise to two populations with different characteristic times. 
Calibration of the detector was done using
neutron source and ambient neutrons, which produce nuclear recoils, 
and gamma-ray source, which produces 
electron recoils via Compton scattering. Nuclear recoils were 
also expected from WIMP-nucleus interactions, whereas electron recoils
from gammas constituted the main background.
Based on the absence of the nuclear recoils in underground data,
the 90\% C.L. on the
number of nuclear recoils was extracted and used to calculate the
limit on the WIMP-nucleon cross-section as a function of WIMP mass. 
The preliminary
limit on the spin-independent WIMP-nucleon cross-section from
293\,kg$\times$days of data is shown in Figure~\ref{limits} in
comparison with the NAIAD limit and some other world-best limits
\cite{nigel}.

\begin{figure}[htb]
\begin{center}
\includegraphics[width=4.0cm,height=6.0cm]{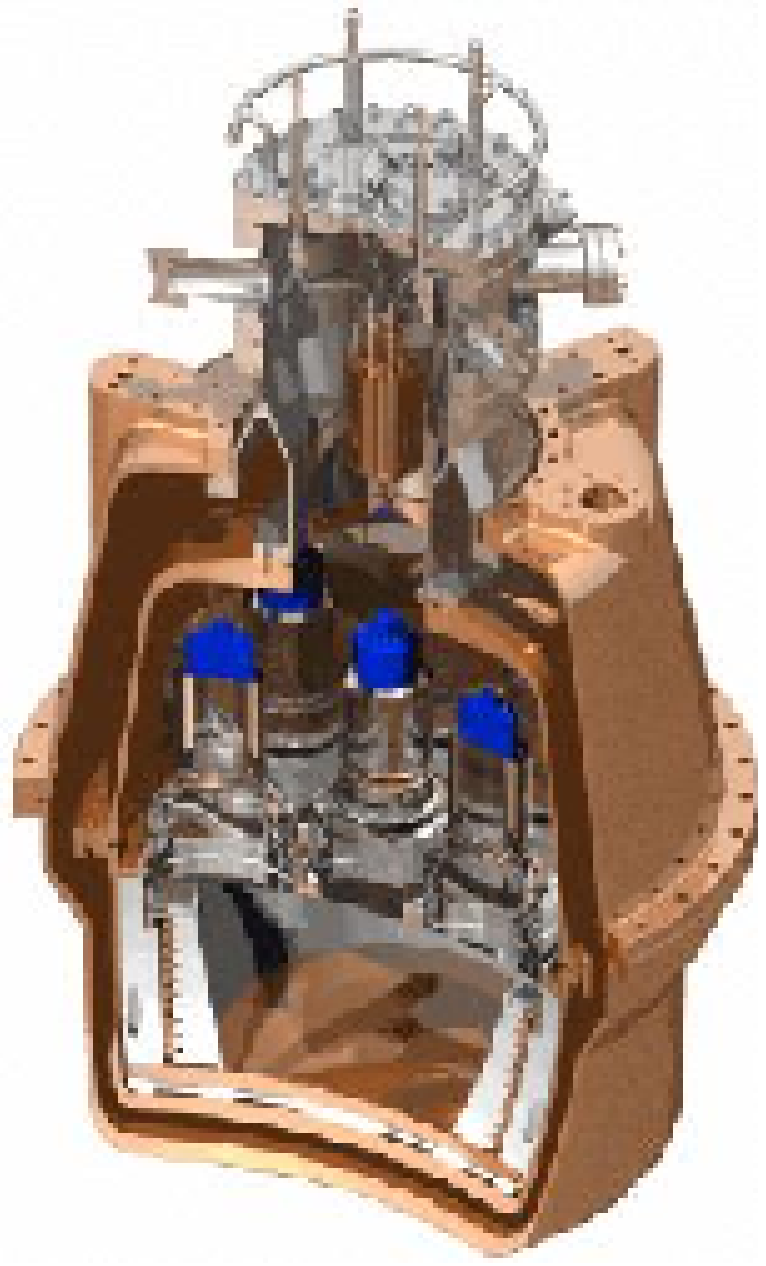}
\label{z2}
\hspace{3.0cm}\includegraphics[width=6.0cm,height=6.0cm]{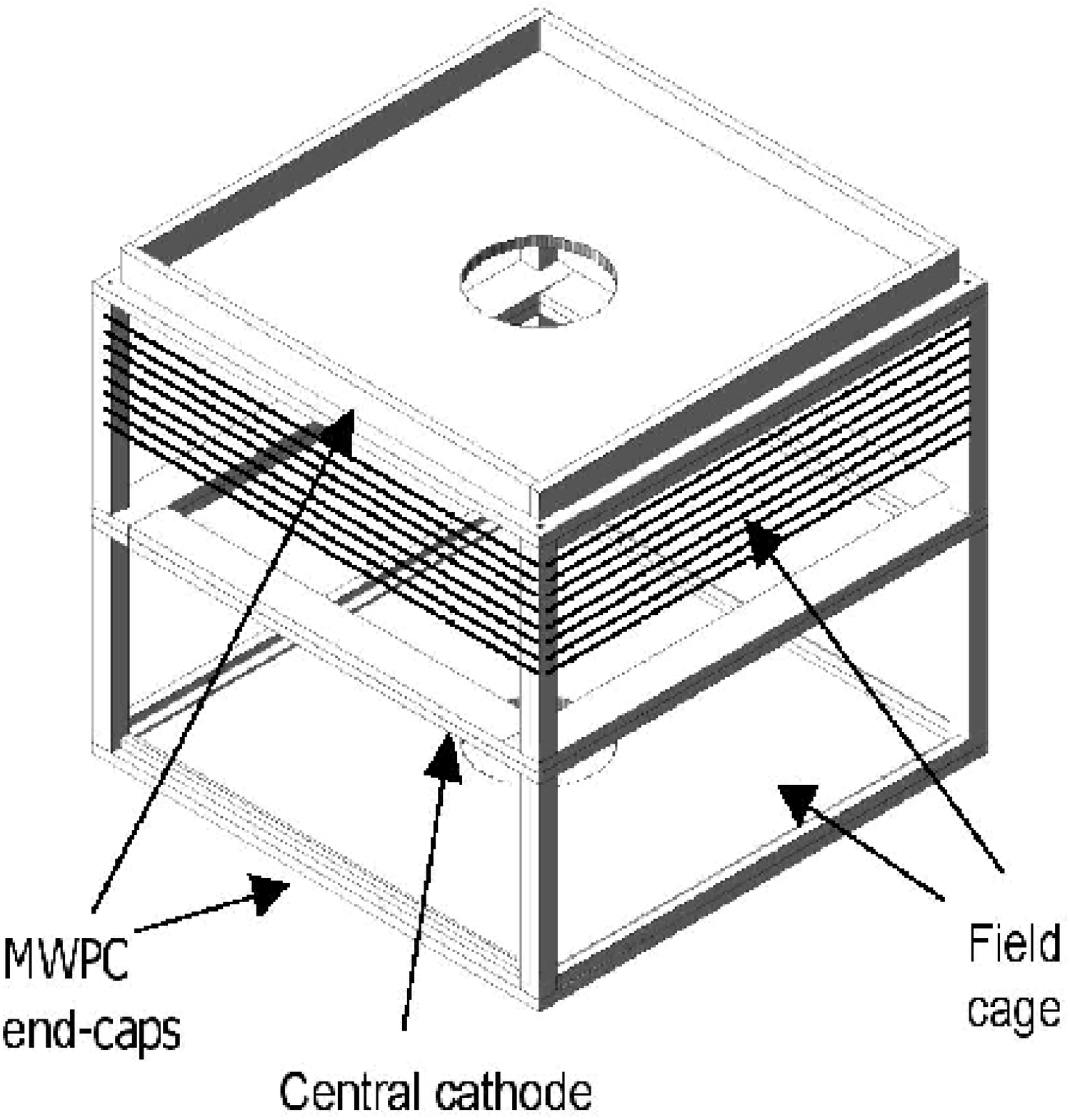}
\label{drift}
\parbox{7.0cm}{Figure 3: Computer view of ZEPLIN II.}
\hspace{0.3cm}
\parbox{8.4cm}{Figure 4: Schematic of the inner part of DRIFT I.}
\end{center}
\end{figure}

Work is now underway on ZEPLIN II and ZEPLIN III detectors. 
ZEPLIN II (Figure ~\ref{z2}) is a two-phase
detector with a target mass of about 30 kg and a sensitivity to WIMP-nucleon
cross-section down to 10$^{-7}$ pb or better at the minimum of the sensitivity 
curve. In ZEPLIN II, recoils
produce both excitation and ionisation in liquid xenon. 
Recombination of electrons and ions produced via ionisation
is prevented by the strong electric field. Electrons, drifting in 
this field towards gas phase, produce a secondary luminescence signal in 
the gas. For a given primary
amplitude an electron recoil will produce a much larger secondary than a
nuclear recoil. This provides ZEPLIN II with greater discrimination power
over ZEPLIN I. ZEPLIN III aims
to increase background discrimination by increasing electric field 
through the liquid xenon and, with a fiducial mass of
6 kg, should achieve similar or better sensitivities to ZEPLIN II.

The ZEPLIN projects described above can be viewed as a development
programme aimed at determining the optimum design for a large-scale
(about one tonne) xenon detector capable of reaching the sensitivity 
below 10$^{-9}$ pb in 1
year of operation. Such a large mass is required to achieve
sufficient signal counts (10 - 100 events). 
ZEPLIN MAX -- a tonne scale liquid xenon experiment -- 
is at the R\&D stage at present.

\vspace{-0.2cm}
\section{The DRIFT Experiment}

The DRIFT (Directional Recoil Identification From Tracks) detector
adopts a different approach to identifying a potential
WIMP signal. DRIFT I uses a low pressure CS$_2$ gas TPC
capable of measuring the components of recoil track
ranges in addition to their energy. 
The use of negative ions, notably CS$_{2}$, to
capture and drift the ionisation electrons reduces diffusion. The 
detector consists of two 0.5 m$^3$
fiducial volumes defined by 0.5 m long field cages mounted either
side of a common cathode plane (Figure~4). 
Particle tracks are read out with two 1 m long MWPCs, one at
either end of the field cages. The difference in track range between
electrons, alpha particles and recoils is such that rejection
efficiencies as high as 99.9\% at 6 keV measured energy 
are possible. After 1 year of
operation DRIFT I is expected to reach a sensitivity of $\sim$10$^{-6}$ pb.
The power of DRIFT comes from its ability to determine the direction
of a WIMP induced nuclear recoil. The Earth's motion around the
galactic centre means that the Earth experiences a WIMP ``wind". As the
Earth rotates through this wind, the nuclear recoil direction is
modulated over a period of one sidereal day, making it a strong
signature of a galactic WIMP signal.

Building on DRIFT I the long term objective of the DRIFT programme is
to scale-up detectors towards a target mass of 100 kg (DRIFT III)
through development of several intermediate scale modules (DRIFT II), 
which can be replicated many times. DRIFT II is
proposed to have 30-50 times the sensitivity of DRIFT I through an
increase in the volume (several modules) and gas pressure. 
A higher gas pressure means
that the recoil range will be shorter requiring higher spatial
resolution. Alternative read-out schemes are currently being
investigated including Gas Electron Multipliers and a MICROMEGAS
microstructure detector.

\vspace{-0.5cm}
\section{Conclusions}

Liquid xenon has been demonstrated as an excellent technology for dark
matter searches with ZEPLIN I already producing significant
sensitivity using pulse shape discrimination. The 
collaboration is now progressing to two phase operation which
shows promise for substantial improvement in sensitivity towards
10$^{-8}$ pb. There is a route with this
technology to reach 10$^{-10}$ pb with an one tonne liquid xenon
detector.
Directional detectors based on low pressure gas provide a unique means
of determining the galactic origin of an observed WIMP signal, a
significant advantage over conventional dark matter experiments.  The
DRIFT programme based on this concept is underway to approach this
possibility, complementing also the liquid xenon programme through the use
of entirely different technology with different target nuclei.
More on the Boulby programme can be found in recent reviews 
\cite{nigel,neil}.

We would like to thank PPARC for financial support. We are also grateful
to the CPL and the Boulby Mine staff for assistance.

%
%

\end{document}